\input{aipcheck}

\documentclass[
    ,final            
  ]
  {aipproc}

\layoutstyle{8x11single}

\newcommand{\inte}{$INTEGRAL$}
\newcommand{\xmm}{$XMM-Newton$}

\newcommand{\sw}{$Swift$}

\def \hcm {\hbox {\ifmmode $ atom cm$^{-2}\else atom cm$^{-2}$\fi}}

\def \ATel {Astron.\ Tel.}
\def \apj {ApJ}
\def \apjl {ApJL}
\def \apjs {ApJS}
\def \aap {A\&A}

\def \pasj {PASJ}

\def \mnras {MNRAS}
\def \iaucirc {IAU Circ.}

\begin{document}

\title{New Galactic High Mass X--ray Binaries discovered with INTEGRAL}


\author{Lara Sidoli}{
  address={INAF-IASF, via Bassini 15, I-20133 Milano, Italy}
}

\begin{abstract}
I will review the main observational 
properties of the new Galactic High Mass X--ray Binaries (HMXBs)
discovered  by the 
$INTEGRAL$ satellite in the hard energy range 17--100 keV. 
About 70\% of the newly discovered HMXBs host OB supergiant companions and show peculiar properties
with respect to classical HMXBs detected with previous missions: some of them display
huge local absorptions, in excess of 10$^{23}$~cm$^{-2}$ (the so-called {\em obscured sources}), while
others show fast transient X--ray emission, leading to the definition of a new sub-class of HMXBs,
the so-called {\em Supergiant Fast X--ray Transients} (SFXTs). 
Their peculiar behavior is still poorly understood and represents a challenge to theory.
\end{abstract}

\maketitle


\section{New Galactic High Mass X--ray Binaries: The role of INTEGRAL discoveries}

Our view of the X--ray sky changed since the launch of the
{\em International Gamma-Ray Astrophysics Laboratory} 
in October 2002 ($INTEGRAL$, \cite{Winkler2003}): 
723 hard X--ray sources have been observed down to a sensitivity limit
at mCrab level \citep{Bird2010} in the energy band 17--100 keV, 
30\% of which are still unidentified.

Among all detected sources, 13\% are High Mass X--ray Binaries (HMXBs).
About 50 objects among the new sources discovered with $INTEGRAL$ ($IGR$ sources), 
have been classified as HMXBs. 
About 70\% of them host supergiant donors
(see also the IGRs on-line list at \url{http://irfu.cea.fr/Sap/IGR-Sources/}), almost
tripling the number of known Galactic HMXBs with supergiant companions.

Among these new HMXBs, two peculiar behaviors have been recognized:

\begin{itemize}

\item the so-called {\em obscured sources}, which display huge amount of low energy absorption
(well in excess of 10$^{23}$~cm$^{-2}$, thought to be local because of its variability, and
produced by the dense wind of the supergiant companion; 
\cite{Walter2003, Walter2006});

\item a new subclass of transients, the {\em Supergiant Fast X--ray Transients} 
(SFXTs; \cite{Sguera2005, Negueruela2005a}). 
Note that a high absorbing column density
is {\em not} a characterizing property of SFXTs, 
although a few SFXTs show high and variable absorption.

\end{itemize}

The discovery of new types of supergiant HMXBs was somehow unexpected:
indeed, before $INTEGRAL$ discoveries, HMXBs hosting OB supergiants were known 
to exhibit  persistent X--ray emission, driven by the accretion
from the strong wind of the blue supergiant companions, mainly in narrow orbits.

They were thought to be easy to detect because of their luminous persistent emission
($\sim$10$^{36}$erg~s$^{-1}$).
The number of Galactic wind-fed HMXBs composed of pulsating neutron stars and 
OB supergiant companions was limited to a few sources
(Vela X--1, 2S0114+650, GX301--2, 1E1145.1--6141, 4U1538--53, X1908+075, XTE~J1855--026).

Then, after the launch of $INTEGRAL$, 
thanks to its sensitivity at hard X--rays and the monitoring strategy of the Galactic
plane, several new HMXBs could
be discovered and identified, 
either highly obscured or with fast transient X--ray emission with long duty cycles,
very difficult to discover with previous missions.

\subsection{Obscured sources:  the prototype IGRJ~16318--4848}

IGRJ16318--4848 is the first source discovered in 2003 by 
the IBIS/ISGRI detector on-board \inte\ \citep{Courvoisier2003}.
The source 
displayed a variable hard X--ray emission on time scales of 1000~s, 
with an average
flux of 6$\times10^{-11}$~erg~cm$^{-2}$~s$^{-1}$ (20--50 keV).

\xmm\ follow-up observations allowed to refine the position \citep{Schartel2003}, and 
revealed an extremely absorbed X--ray spectrum,
with an absorbing column density of 2$\times$10$^{24}$~cm$^{-2}$, together with prominent
iron K$_{\alpha}$ and K$_{\beta}$, and nichel K$_{\alpha}$ emission lines  
(\cite{dePlaa2003}, \cite{Matt2003}, \cite{Walter2003}, \cite{Ibarra2007}), suggestive
of the presence of a dense local envelope 
(\cite{Matt2003}, \cite{Revnivtsev2003}, \cite{Kuulkers2005}).

X--ray observations with $Suzaku$ \citep{Barragan2009} confirmed the X--ray spectrum 
rich in iron and nichel emission lines and allowed a truly simultaneous broad band spectroscopy (1--60 keV).
These authors confirmed the huge amount of absorption, but found a significantly harder X--ray spectrum,
well described with a flat power law with a 
photon index of $\sim$0.67 (assuming an absorbed exponentially cutoff
power law, together with Gaussian emission lines).

IGRJ16318--4848 is optically associated with a sgB[e] star, with a highly uncertain
distance (0.9--6.2 kpc), implying an X--ray 
luminosity in the range between 1.3$\times$$10^{35}$~erg~s$^{-1}$ and 
6.2$\times$$10^{36}$~erg~s$^{-1}$ (2--10 keV). 
If it is located in the Norma Arm,
it lies at a distance of about 4.8 kpc \citep{Filliatre2004}.
The optical to MIR spectral energy distribution suggests the presence of a mid-infrared excess
above the stellar spectrum, indicative of dust. This findings, together with the high absorption,
imply dust and dense cool gas enshrouding the whole binary system 
(\cite{Rahoui2008}, \cite{Chaty2008}).

The B[e]-phenomenon is peculiar of very different kinds of stars (from supergiants to pre-main sequence stars) 
and is characterized by spectra with a near infrared excess (attributed to emission
from warm dust) and by  forbidden emission lines (e.g. Fe~II)
\cite{Lamers1998}. In particular, B[e] supergiants (sgB[e], \cite{Zickgraf1985, Lamers1998}) 
are thought to be in an evolutionary stage
intermediate between OB supergiant and Wolf Rayet star \citep{Clark1999}.
Their properties have been explained in terms of 
a two component stellar wind, with the presence of a circumstellar outflowing disk wind \citep{Zickgraf1985}.

After IGRJ16318--4848, \inte\ discovered other similar sources  displaying 
high local absorption at soft X--rays (larger than 10$^{23}$~cm$^{-2}$)
associated with blue supergiant companions. 
Thus, IGRJ16318--4848 became the prototype of the so-called
{\em obscured} HMXBs, being the first and most extreme example of this sub-class of massive X--ray binaries 
(e.g. \cite{Kuulkers2005, Chaty2005}). It is also one of the rare examples of HMXB with a companion
showing the B[e] phenomenon \citep{Clark1999}.

\subsection{Supergiant Fast X--ray Transients}

Supergiant Fast X--ray Transients (SFXTs) are a sub-class
of massive X--ray binaries recognized mainly thanks to \inte\ discoveries \cite{Sguera2005, Negueruela2005a}.

SFXTs are characterized by transient and fast X--ray emission (composed by short and bright 
flares, with a duration of a few hours, as observed with \inte,  \cite{Sguera2005}, \cite{Negueruela2005a}) 
and by the association with OB supergiant companions 
(e.g. \cite{Pellizza2006}, \cite{Masetti2006}, \cite{Negueruela2006}, \cite{Rahoui2008}, \cite{Nespoli2008}).
These two main properties have led to the identification 
of a new class of 9 members to date 
(together with several
other SFXTs candidates, with fast transient X--ray emission, 
but still unknown counterparts).
Optical/IR spectroscopy allowed the determination of the source distance, 
implying an X--ray luminosity
at the flare peak of 10$^{36}$--10$^{37}$~erg~s$^{-1}$.

$INTEGRAL$ is able to observe the brightest flares, with 
no persistent emission outside flaring activity.
More sensitive instruments allow to better investigate the 
properties of SFXTs when they are not undergoing
an outburst.

The quiescence, characterized by a very soft (likely thermal) spectrum, has been rarely 
observed in SFXTs. It is at a luminosity level of 
$\sim$10$^{32}$~erg~s$^{-1}$, as observed in IGR~J17544--2619 with $Chandra$ \citep{zand2005}.
In literature, the term ``quiescence'' is often used with different meanings, sometimes simply
to indicate the X--ray emission when the source is not undergoing a bright outburst 
(``out-of-outburst'' emission), 
whereas we mean here a specific source state at 10$^{32}$~erg~s$^{-1}$,
where no accretion onto the compact object 
is present and the spectrum is very soft and thermal \citep{zand2005}. 
This quiescent luminosity implies a very high dynamic range (ratio between the peak luminosity and the luminosity 
during quiescence) of 4-5 orders of magnitude.

Observations with \xmm\ caught a few SFXTs in a very low level of emission (L${_X}$$\sim$10$^{32}$~erg~s$^{-1}$),
showing a hard 0.5--10 keV spectrum and faint very short flares, 
suggestive of some level of accretion onto the compact object
\citep{Gonzalez2004, Bozzo2010}.

The presence of a very low level of accretion 
is also suggested by our 
recent observation of the SFXT IGR\,J08408--4503 
with $Suzaku$ \citep{Sidoli2010:igr08408} performed in December 2009, which 
caught the source in an initial 
low intensity state at 4$\times$10$^{32}$~erg~s$^{-1}$4 (0.5--10 keV),
during the first 120 ks. Then, the source showed two long (lasting about 45 ks each) 
and faint flares.
The spectrum of both the pre-flare and flare emission resulted in a hard
X--ray emission, well fitted by a double component spectrum, 
with a soft thermal plasma model (kT$\sim$0.2--0.3 keV) 
together with a power law (photon index $\sim$ 2), 
differently absorbed.
The long faint flares are probably produced by the direct 
accretion of large structures in the 
supergiant wind (similar to corotating interaction regions, \cite{Mullan1984}), 
with an estimated thickness of 10$^{12}$~cm \citep{Sidoli2010:igr08408}.

\begin{figure}
  \includegraphics[height=.3\textheight]{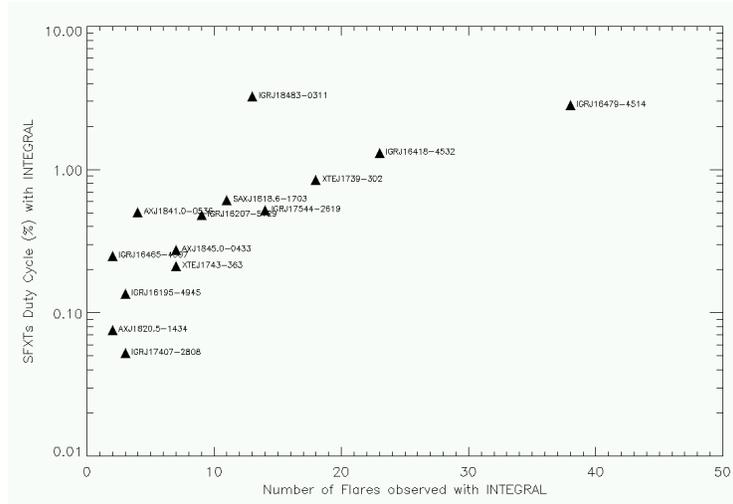}
  \caption{INTEGRAL results on SFXTs: duty cycles 
(percentage of time spent in bright flares with respect
to the total observing time with \inte\ of the source position) calculated from 
data reported by \cite{Ducci2010}. The sources are both confirmed and candidate SFXTs
observed during the survey of the central region of our Galaxy (see \cite{Ducci2010} for details).}
\label{lsidoli:fig:dc}
\end{figure}

\begin{figure}
  \includegraphics[height=.35\textheight]{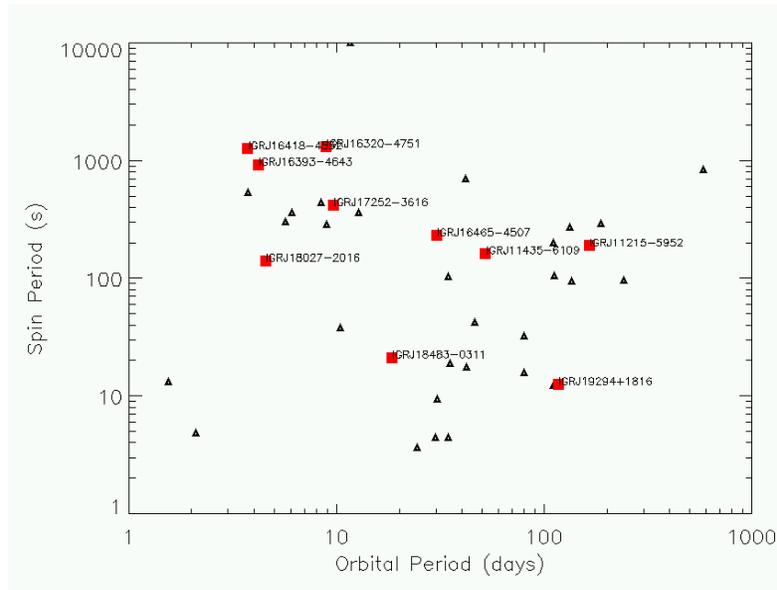}
  \caption{New Galactic X--ray binaries discovered by \inte\ (red squares and source names)
in the Corbet diagram of 
spin period versus orbital period. 
The solid triangles indicate previously known Galactic 
HMXBs and Be/X--ray transients. 
Note that almost all the new Galactic IGRs reported here 
are HMXBs hosting supergiant companions,
except IGR~J16393--4643 (likely a Symbiotic system) and  IGR J11435--6109 (Be star).
The nature of the donor star in IGR~J19294+1816 (main sequence Be or OB supergiant) 
is still unclear \citep{Rodriguez2009}.
}
\label{lsidoli:fig:corbet}
\end{figure}

%
%
A monitoring with Swift/XRT of 4 members of the class, spanning about two years of observations 
(starting in October 2006, \cite{Sidoli2008:sfxts_paperI, Romano2010}),
demonstrated that the quiescence is a rare state for these transients.
Indeed, most of their lifetime is spent in an intermediate state of emission
in the range  between 10$^{33}$ and 10$^{34}$~erg~s$^{-1}$ (0.3--10 keV), with a hard spectrum
(well described by a hard power law with photon index of $\sim$1--2 or hot  black body with 
temperature kT$\sim$1-2 keV) and high level of flux variability. 
The spectral hardness and the highly variable X--ray 
intensity indicate that the sources are in accretion even outside the bright outbursts.

An important property is their spectral similarity with accreting X--ray pulsars:
the broad-band spectrum (0.5--100 keV) during bright flares is well fitted by an absorbed 
hard power law (photon index $\sim$1) with a high energy cutoff at $\sim$10--30 keV, or with Comptonized
emission \citep{SidoliPM2006, Gotz2007:08408-4503, Romano2008:sfxts_paperII, 
Sidoli2009:sfxts_paperIII, Sidoli2009:sfxts_paperIV, Romano2009:sfxts_paper08408}.
The absorbing column density is often variable \citep{Sidoli2009}, indicating
that it is local and close to the X--ray source.

The SFXTs outbursts last a few days
(\cite{Romano2007, Sidoli2008:sfxts_paperIII, Rampy2009:suzaku17544}), 
as demonstrated for the 
first time by an X--ray monitoring 
of an outburst of the SFXT IGRJ~11215--5952 in February 2007 (\cite{Romano2007, Sidoli2007})
and later confirmed during the \sw/XRT monitoring of other 4 SFXTs during their outbursts  \citep{Sidoli2009}.
Each ouburst is actually composed by an enhanced average X--ray emission together with several bright
flares lasting a few hours (\cite{Romano2007, Rampy2009:suzaku17544}).

Source duty cycles are small, 
although highly variable from source to source: during the Swift/XRT 
monitoring, the time spent in bright outburst is 3\%-5\% in 3 SFXTs, \cite{Romano2010},
while with \inte\ it is much smaller (see Fig.~\ref{lsidoli:fig:dc}). 
The figure has been obtained from data reported by \cite{Ducci2010} 
and includes both confirmed and candidate SFXTs after seven
years of \inte\ operations.

\begin{table}
\begin{tabular}{lccl}
\hline
 \tablehead{1}{r}{b}{Source \\ }
  & \tablehead{1}{r}{b}{Orbital Period \\ (d) }
  & \tablehead{1}{r}{b}{Spin Period \\ (s) }
   & \tablehead{1}{r}{b}{References \\  } \\
\hline
IGRJ~08408--4503 & 35 (?)               & -                    & P$_{orb}$: \cite{Romano2009:sfxts_paper08408}  \\
IGRJ~11215--5952 & 164.6                & 186.78$\pm{0.3}$     & P$_{orb}$: \cite{SidoliPM2006, Romano2009:11215_2008}; 
P$_{spin}$: \cite{Swank2007:atel999, Sidoli2007}  \\
IGRJ~16465--4507 & 30.243$\pm{0.035}$   & 228$\pm{6}$          & P$_{orb}$: \cite{Laparola2010, Clark2009}; P$_{spin}$: \cite{Lutovinov2005} \\
IGRJ~16479--4514 & 3.3194$\pm{0.0010}$  & -                    & P$_{orb}$: \cite{Jain2009:16479} \\
XTE~J1739--302   & 51.47$\pm{0.02}$     & -                    & P$_{orb}$: \cite{Drave2010} \\
IGRJ~17544--2619 & 4.926$\pm{0.001}$    & -                    & P$_{orb}$: \cite{Clark2010} \\
SAXJ~1818.6--1703& 30.0$\pm{0.1}$       & -                    & P$_{orb}$: \cite{Zurita2009, Bird2009} \\
AX~J1841.0--0536 & -                & 4.7394$\pm{0.0008}$  & P$_{spin}$: \cite{Bamba2001} \\
IGR~J18483--0311 & 18.55 $\pm{0.03}$    & 21.0526$\pm{0.0005}$ &  P$_{orb}$: \cite{Levine2006};    P$_{spin}$:  \cite{Sguera2007}  \\
\hline
\end{tabular}
\caption{List of Supergiant Fast X--ray Transients }
\label{lsidoli:tab:sfxt}
\end{table}

A few SFXTs are X--ray pulsars (see Table~\ref{lsidoli:tab:sfxt} and Fig.~\ref{lsidoli:fig:corbet}), 
thus demonstrating that
the compact object is a neutron star,
while in the other sources a black hole cannot be excluded, although the spectral similarity
to accreting pulsars is suggestive of a neutron star.
The SFXTs spin periods are in the range from 4.7~s to 228~s. 
The orbital periods are also very different, between 3.3 days and 165 days (Table~\ref{lsidoli:tab:sfxt}).
Most of them have been determined from the modulation of the X--ray light curve, while
in the case of IGRJ~11215--5952 \cite{SidoliPM2006} it has been derived from the periodically
recurrent outbursts.
The orbital period of $\sim$35~days in  IGRJ~08408--4503 (Table~\ref{lsidoli:tab:sfxt}) 
has been suggested by \cite{Romano2009:sfxts_paper08408}
based on the  duration of the flares, which is thought to be linked to the orbital separation, in the framework
of  the new wind accretion model
of \cite{Ducci2009}. Thus, it needs to be confirmed by timing analysis.

Interestingly, a few of the newly discovered HMXBs hosting supergiants 
lie in the region typical of Be/X--ray transients (IGRJ~11215--5952, IGR~J19294+1816, IGRJ~18483--0311) 
in the Corbet diagram  (Fig.~\ref{lsidoli:fig:corbet}), 
probably suggesting a possible evolutionary link with these transients (see, e.g., \cite{Liu2010} 
and Chaty 2010, these proceedings).

\subsection{The mystery of SFXTs}

SFXTs are massive binaries where the X--ray emission is thought to be produced by the
direct accretion of the supergiant wind onto the compact object, although the possible
formation of transient accretion disks have been suggested \cite{Ducci2010}.
On the other hand, the wind accretion alone cannot explain a so different behavior 
in two sub-classes of HMXBs (SFXTs vs persistent HMXBs with supergiant donors)
where the donor star, the compact object (neutron star), the orbital parameters seem to be very similar. 
SFXTs like IGRJ~16479--4514 \citep{Jain2009:16479} 
and IGRJ~17544--2619 \citep{Zurita2009, Bird2009} are in narrow orbits, even
narrower than several persistent HMXBs, thus ruling out the hypothesis 
of wide eccentric orbits \citep{Negueruela2008} 
as the 
main property which makes the difference between these SFXTs and persistent 
HMXBs containing supergiants. 

Different mechanisms have been proposed to explain SFXTs (see \cite{Sidoli2009:cospar} for a review):
accretion from clumpy winds, where the SFXTs flares are produced by the sudden accretion of a dense clump
\cite{zand2005, Walter2007} in a binary system with a wide and possibly eccentric orbit \citep{Negueruela2008}; 
accretion from aspherical winds (with a preferential plane for the outflowing wind), where
the flares are triggered by the neutron star passage inside this outflow \cite{Sidoli2007}; 
centrifugal or magnetic barriers which halt the accretion onto the neutron star for most of the time \cite{Grebenev2007, Bozzo2008}.
Each these mechanisms can hardly  explain the behavior of {\em all} SFXTs, whose link with persistent HMXBs remain unclear.
It is also possible that different mechanisms are at work together in a single SFXT, 
or, alternatively, that SFXTs is a non-homogeneous class. 
Thus, despite the large amount of observational data, there are still several open issues that need to be addressed:
the accretion mechanism, the link between different kinds of HMXBs, the SFXTs evolutionary path and formation history.


\begin{theacknowledgments}
This work was supported in Italy by contract ASI/INAF I/088/06/0.
\end{theacknowledgments}



\bibliographystyle{aipproc}   


\IfFileExists{\jobname.bbl}{}
 {\typeout{}
  \typeout{******************************************}
  \typeout{** Please run "bibtex \jobname" to optain}
  \typeout{** the bibliography and then re-run LaTeX}
  \typeout{** twice to fix the references!}
  \typeout{******************************************}
  \typeout{}
 }

\end{document}